\documentclass[preprint,3p,twocolumn]{elsarticle} 
\journal{Astroparticle Physics}
\usepackage{amsmath,amssymb,booktabs,graphicx,verbatim}
\usepackage[colorlinks=true,citecolor=black,filecolor=black,linkcolor=black,urlcolor=black,pdftex]{hyperref}
\newcommand{\1}[1]{\, \mathrm{#1}} 
\newcommand{\n}[1]{\mathrm{#1}}    
\newcommand{\percent}{\%}

\newcommand{\order}{\mathcal{O}}
\newcommand{\arxiv}[1]{\href{http://arxiv.org/abs/#1}{\texttt{arXiv:#1}}}
\begin{document}

\begin{frontmatter}
\title{Electron and Gamma Background in CRESST Detectors}

\author[mp]{R.~F.~Lang}
\ead{rafael.lang@mpp.mpg.de}
\author[mp]{G.~Angloher}
\author[tb]{M.~Bauer}
\author[mp]{I.~Bavykina}
\author[mp,co]{A.~Bento}
\author[ox]{A.~Brown}
\author[gs]{C.~Bucci}
\author[tu]{C.~Ciemniak}
\author[tu]{C.~Coppi}
\author[tb]{G.~Deuter}
\author[tu]{F.~von~Feilitzsch}
\author[mp]{D.~Hauff}
\author[ox]{S.~Henry}
\author[mp]{P.~Huff}
\author[ox]{J.~Imber}
\author[ox]{S.~Ingleby}
\author[tu]{C.~Isaila}
\author[tb]{J.~Jochum}
\author[mp]{M.~Kiefer}
\author[tb]{M.~Kimmerle}
\author[ox]{H.~Kraus}
\author[tu]{J.-C.~Lanfranchi}
\author[mp,ox]{B.~Majorovits}
\author[ox]{M.~Malek}
\author[ox]{R.~McGowan}
\author[ox]{V.~B.~Mikhailik}
\author[mp]{E.~Pantic}
\author[mp]{F.~Petricca}
\author[tu]{S.~Pfister}
\author[tu]{W.~Potzel}
\author[mp]{F.~Pr\"obst}
\author[tu]{S.~Roth}
\author[tb]{K.~Rottler}
\author[tb]{C.~Sailer}
\author[mp]{K.~Sch\"affner}
\author[mp]{J.~Schmaler}
\author[tb]{S.~Scholl}
\author[mp]{W.~Seidel}
\author[mp]{L.~Stodolsky}
\author[ox]{A.~J.~B.~Tolhurst}
\author[tb]{I.~Usherov}
\author[tu,de]{W.~Westphal}

\address[mp]{Max-Planck-Institut f\"ur Physik, F\"ohringer Ring 6,
D-80805 M\"unchen, Germany}
\address[tu]{Physik-Department E15, Technische Universit\"at
M\"unchen, D-85747 Garching, Germany}
\address[ox]{Department of Physics, University of Oxford, Oxford
OX1 3RH, United Kingdom}
\address[tb]{Eberhard-Karls-Universit\"at T\"ubingen, D-72076
T\"ubingen, Germany}
\address[gs]{INFN, Laboratori Nazionali del Gran Sasso, I-67010
Assergi, Italy}
\address[co]{on leave from: Departamento de Fisica, Universidade de
Coimbra, P3004 516 Coimbra, Portugal}
\address[de]{Deceased}

\begin{abstract}
The CRESST experiment monitors $300\1{g}$ $\n{CaWO_4}$ crystals as targets for particle interactions in an ultra low background environment. In this paper, we analyze the background spectra that are recorded by three detectors over many weeks of data taking. Understanding these spectra is mandatory if one wants to further reduce the background level, and allows us to cross-check the calibration of the detectors. We identify a variety of sources, such as intrinsic contaminations due to primordial radioisotopes and cosmogenic activation of the target material. In particular, we detect a $3.6\1{keV}$ X-ray line from the decay of $\n{{}^{41}Ca}$ with an activity of $(26\pm4)\1{\mu Bq}$, corresponding to a ratio $\n{{}^{41}Ca}/\n{{}^{40}Ca}=(2.2\pm0.3)\times10^{-16}$.
\end{abstract}


\end{frontmatter}

\section{Introduction}

The Cryogenic Rare Event Search with Superconducting Thermometers CRESST~\cite{angloher2005,angloher2009} aims to detect rare nuclear recoils from elastic scattering of Dark Matter particles~\cite{jungman1996}. Since less than one such recoil is expected per kilogram of target mass and week of exposure, efficient shielding of the target against ambient radioactivity is mandatory. 

To this end, the CRESST experiment is located in the Laboratori Nazionali del Gran Sasso under an average of $1400\1{m}$ of rock overburden, where the cosmic muon flux is reduced by about six orders of magnitude with respect to the surface. Additionally, the target is surrounded (from the outside to the inside) by $45\1{cm}$ of polyethylene to reduce the neutron flux, a radon box that is constantly flushed with nitrogen gas to reduce the concentration of radioactive radon and its daughters inside the shielding, as well as $20\1{cm}$ of lead and $14\1{cm}$ of pure copper against the external gamma background~\cite{angloher2009}. The understanding of the remaining radioactive background present in the experiment demonstrates its excellent performance and is a necessary step to improve the sensitivity of CRESST to rare processes.

\section{Setup}

The target consists of individual $300\1{g}$ $\n{CaWO_4}$ crystals that are operated as calorimeters~\cite{angloher2005}. Up to 33 such detectors can be arranged in a compact layout~\cite{angloher2009}. To enhance the temperature signal following an interaction, the crystals are cooled to $\sim 15\1{mK}$ where heat capacities are low. The high frequency phonons that are created in interactions in the crystal are collected by a thermometer, which is a thin tungsten film that is evaporated directly onto the crystals. This film becomes superconducting at these temperatures and is stabilized in its transition to the superconducting state. A small temperature rise results in an increase of resistance that is measured with a SQUID-based readout circuit~\cite{seidel1990}.

A resistive heater structure is evaporated onto each thermometer. This structure is used to provide a constant heating bias as well as for periodic injection of additional heat pulses. Each such injected heat pulse briefly drives the thermometer out of its transition. The amplitude response of the thermometer to the injected heat is controlled to stay at a constant value with small adjustments to the heating bias by means of a feedback system. In this way the operating point of the thermometer is stabilized over several months to within the required precision~\cite{angloher2005,angloher2009}. This can also be seen from the sharpness of gamma lines discussed here.

\section{Pulse Height Evaluation}

Our thermometers yield rather slow pulses with rise times of about $1\1{ms}$. These pulses are digitized with a time base of $40\1{\mu s}$. $4096$ such samples comprise a record, of which the first $1024$~samples are taken from the time before the trigger to define the baseline of the pulse. 

Template pulses are fitted to the record in order to evaluate the amplitude of the pulse as a measure of the deposited energy. These templates are an average of appropriate pulses, often obtained from a gamma calibration line. Since the SQUIDs only measure relative changes in current, the baseline needs to be defined individually for each recorded pulse. Hence, the free parameters in the fit are the level of the baseline, the onset of the pulse, and its amplitude.

The pulse samples the transition curve of the thermometer from its superconducting to the normal conducting state. For interactions with recoil energies below $\sim200\1{keV}$, the resultant pulse generally samples only the linear region of the thermometer's transition curve between the superconducting and normal conducting state. For interactions with larger energies, the pulse samples a region of the transition curve that is non-linear, distorting the resulting pulse shape. Therefore, in the template pulse fit, only those parts of the pulse are included that show a linear behavior. This linearizes the energy response of the fitted amplitudes over a wide energy range, even up to the MeV range~\cite{cozzini2004}.

\section{Energy Calibration}

To reach the cryogenic temperatures, the detectors are enclosed by a fivefold thermal shield made from copper. Calibration sources are placed outside this shield, which has a total thickness of $12\1{mm}$ of copper. This requires gamma energies $\gtrsim 100\1{keV}$ in order to penetrate this barrier. Hence, a method is required to take the energy calibration to lower energies.

To this end, additional heater pulses corresponding to a number of fixed energies are injected into the heater structure every 30 seconds. The pulse height of the response pulse from the thermometer is also evaluated with a template fit. Non-linearities are due to the shape of the transition curve and due to the readout circuit. They manifest themselves in the amplitude of the response pulse, but, for given amplitude, not in its shape. Therefore, the response pulses give the correct energy assessment even if there are small differences in the pulse shape of heater response pulse and particle pulse.

\begin{figure}[htb]
\begin{center}\includegraphics[angle=90,width=1\columnwidth,clip,trim=0 0 130 0]{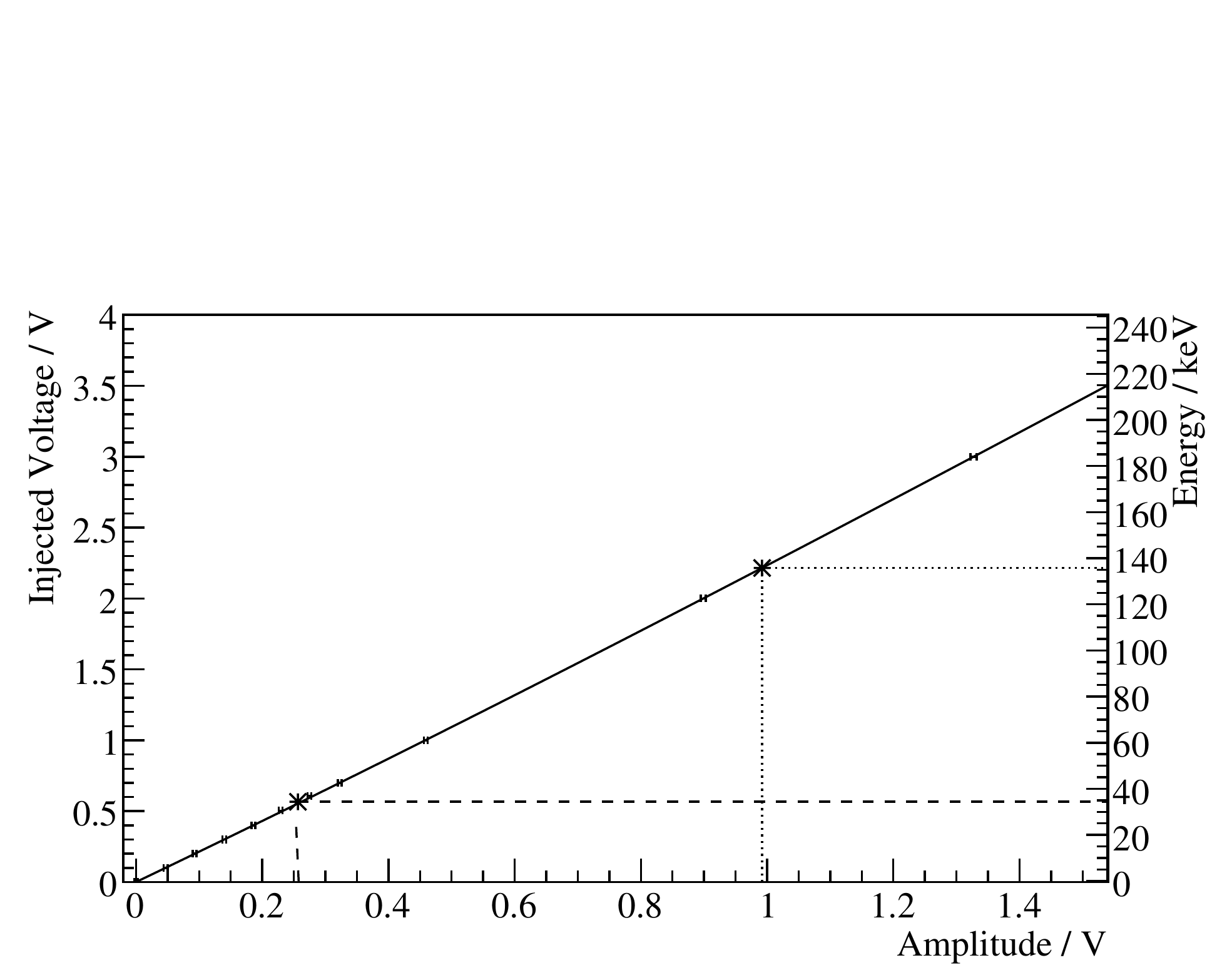}\end{center}
\vspace{-5mm}
\caption{Calibration at low energies with heater pulses of various discrete energies. Fitting a low order polynomial to the thermometer response (small points, including error bars) gives the transfer function (solid line) between injected voltage (scale on the left) and fitted amplitude (scale on the bottom). The $136\1{keV}$ peak gives a calibration point (dotted line) for the injected energy (scale on the right). Then, for each particle pulse, the particle's energy is found by evaluating the transfer function at the fitted amplitude (dashed line).}
\label{fig:cpeb}\end{figure}

The heater pulses are calibrated at relatively high energies, typically with the $122\1{keV}$ or $136\1{keV}$ line from a $\n{{}^{57}Co}$ source. The amplitude response of the thermometer as a function of energy injected into the heater can then be fitted with a low-order polynomial function. Figure~\ref{fig:cpeb} depicts this procedure for the calibration of particle pulses in the complete energy range covered by the heater pulses. In the following we identify a variety of lines at the expected energies, thus validating our calibration procedure.

\section{Calibration Spectrum}\label{sec:copperfluorescence}

\begin{figure*}[htb]
\begin{center}\includegraphics[angle=90,width=1\textwidth,clip,trim=0 0 70 0]{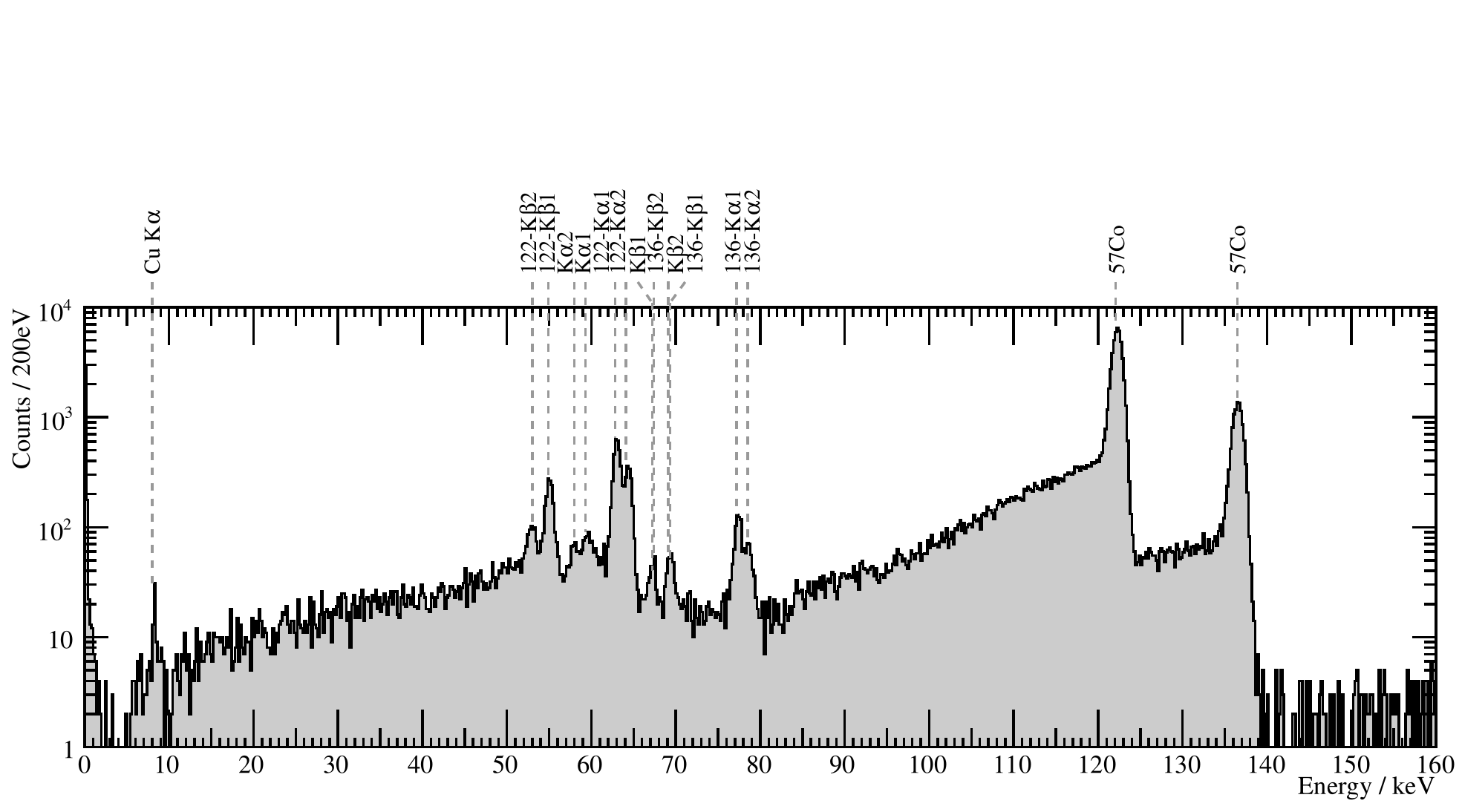}\end{center}
\vspace{-5mm}
\caption{Energy spectrum from a high statistics $\n{{}^{57}Co}$ calibration (detector \textsc{Daisy}, run~31).}
\label{fig:CoSpectrumDaisyRun31g}
\end{figure*}

Figure~\ref{fig:CoSpectrumDaisyRun31g} shows a typical spectrum from a $\n{{}^{57}Co}$ calibration. The two strong peaks at $122\1{keV}$ and $136\1{keV}$ originate from gamma transitions in the $\n{{}^{57}Co}$ source. They show pronounced shoulders towards lower energies, which originate from photons that are Compton scattered in the copper thermal shields and the detector support structures that surround the detectors. 

A gamma can interact in the crystal via the photoelectric effect, ejecting an electron from its shell. The X-ray produced in the subsequent recombination may escape from the crystal. In this case, we measure the energy of the original gamma minus the energy of the X-ray (escape peak). If the escaping X-ray reaches a neighboring crystal, its energy will be measured there. In figure~\ref{fig:CoSpectrumDaisyRun31g} we observe a set of peaks in the energy range between $50\1{keV}$ and $80\1{keV}$, originating from the dominant $K_{\n{\alpha1}}$, $K_{\n{\alpha2}}$, $K_{\n{\beta1}}$/$K_{\n{\beta3}}$ and $K_{\n{\beta2}}$ escape lines of tungsten X-rays. The origin of the various observed lines are labeled in the figure. In addition, a line is visible at $8\1{keV}$ from copper fluorescence due to interactions in the copper surrounding the cryostat (see section~\ref{sec:copper}).

The calibration spectrum allows confirmation of our energy calibration method. This is shown in table~\ref{tab:ecal}, where measured energies are compared to the literature values~\cite{firestone1996} for a few lines. The values agree within the resolution of our detectors.

\begin{table}[htb]
\centering\resizebox{\columnwidth}{!} {\begin{tabular}{lrrr}
\toprule
	line   & literature & reconstructed & resolution	\\
	origin & value      & energy        & ($1\sigma$) \\
\midrule
	Cu $\n{K_{\alpha}}$          & $\phantom{00}8.0\1{keV}$ & $\phantom{00}8.3\1{keV}$ &	$0.1\1{keV}$ \\
	$122\1{keV}-\n{K_{\beta1}}$  & $\phantom{0}54.9\1{keV}$ & $\phantom{0}55.0\1{keV}$ &	$0.4\1{keV}$ \\
	W $\n{K_{\alpha1}}$          & $\phantom{0}59.3\1{keV}$ & $\phantom{0}59.6\1{keV}$ &	$0.5\1{keV}$ \\
	$122\1{keV}-\n{K_{\alpha1}}$ & $\phantom{0}62.8\1{keV}$ & $\phantom{0}63.0\1{keV}$ &	$0.3\1{keV}$ \\
	$\n{{}^{57}Co}$              &           $122.1\1{keV}$ &           $122.2\1{keV}$ &	$0.6\1{keV}$ \\
	$\n{{}^{57}Co}$              &           $136.5\1{keV}$ &           $136.5\1{keV}$ &	$0.6\1{keV}$ \\
\bottomrule
\end{tabular}}\caption{Literature values, reconstructed energies and resolution of a few lines observed in the cobalt calibration. All values agree and validate our energy calibration procedure. The energy resolution is seen to be better than $1\1{keV}$.}\label{tab:ecal}\end{table}

\section{Background Spectra}

\begin{figure}[htb]
\begin{center}\includegraphics[angle=90,width=1\columnwidth,clip,trim=0 0 115 0]{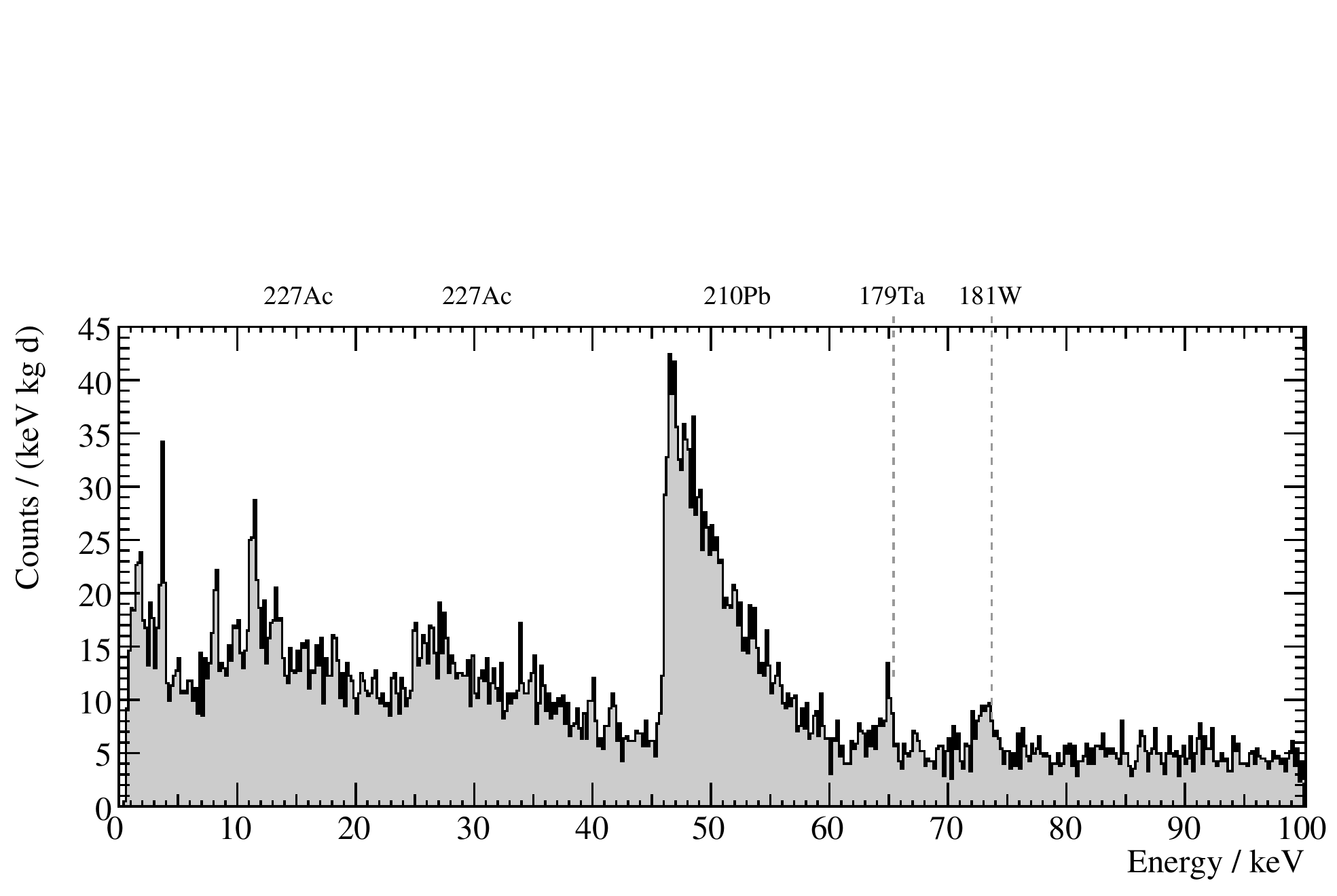}\end{center}
\vspace{-5mm}
\caption{Background spectrum of a second crystal ($m=305.5\1{g}$) from an exposure of $21.20\1{kg\,d}$ in 2007 (detector \textsc{Verena}, run~30).}
\label{fig:verenarun30c}\end{figure}

\begin{figure}[htb]
\begin{center}\includegraphics[angle=90,width=1\columnwidth,clip,trim=0 0 115 0]{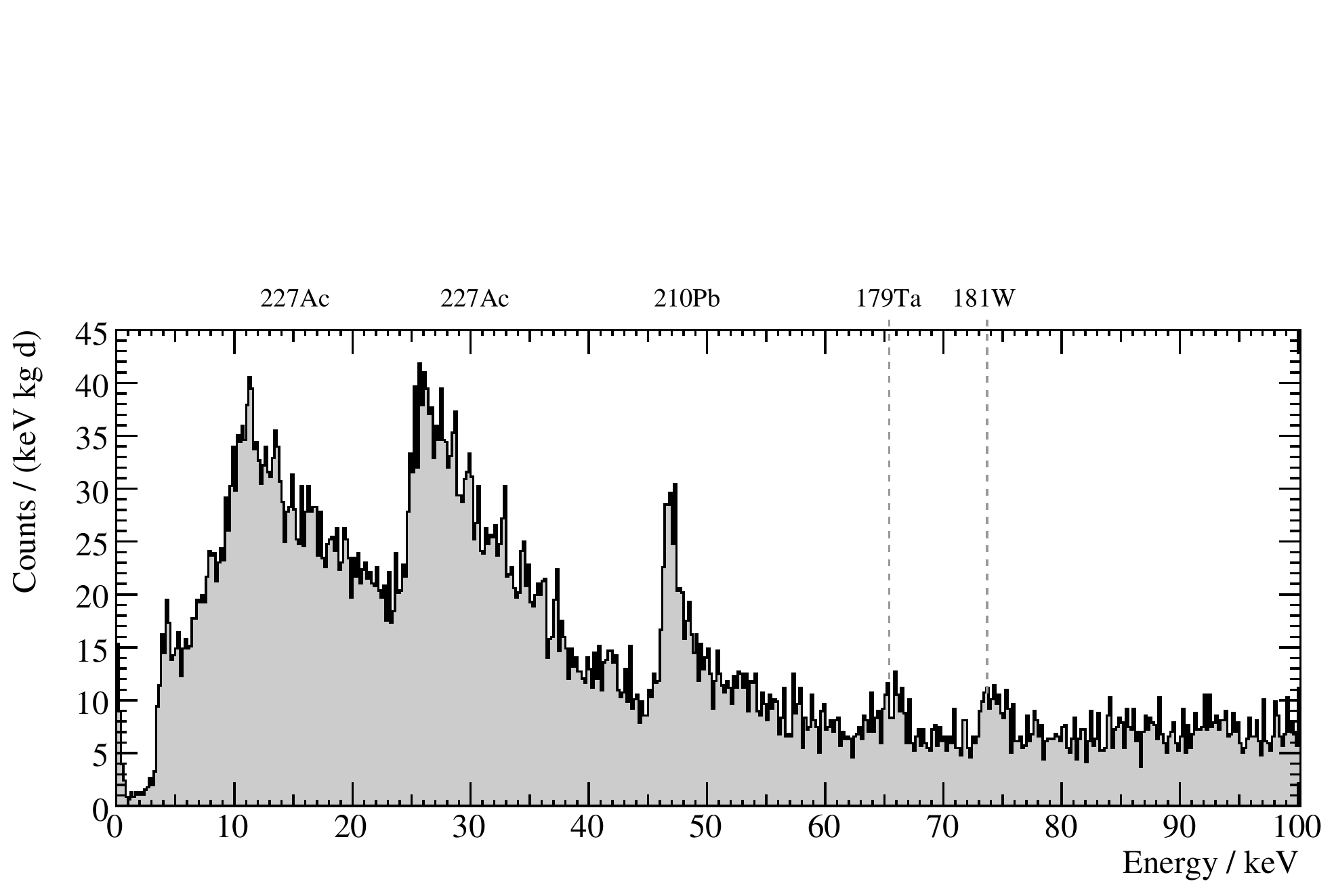}\end{center}
\vspace{-5mm}
\caption{Background spectrum recorded in 2007 from a third detector ($m=307.0\1{g}$) with an exposure of $22.81\1{kg\,d}$ (detector \textsc{Zora}, run~30).}
\label{fig:zorarun30c}\end{figure}

In the following, spectra from exposure to background radiation alone are analyzed, from one detector during the prototyping phase of the experiment~\cite{angloher2005} (figures~\ref{fig:daisystrontium}, \ref{fig:daisyrun28c}) and two detectors during the commissioning phase~\cite{angloher2009} (figures~\ref{fig:verenarun30c}, \ref{fig:verenarun30m} and~\ref{fig:zorarun30c}). All these crystals were grown by the General Physics Institute, Moscow, Russia, but in different batches.

\subsection{Strontium and Yttrium}

\begin{figure}[htb]
\begin{center}\includegraphics[angle=90,width=1\columnwidth,clip,trim=0 0 132 0]{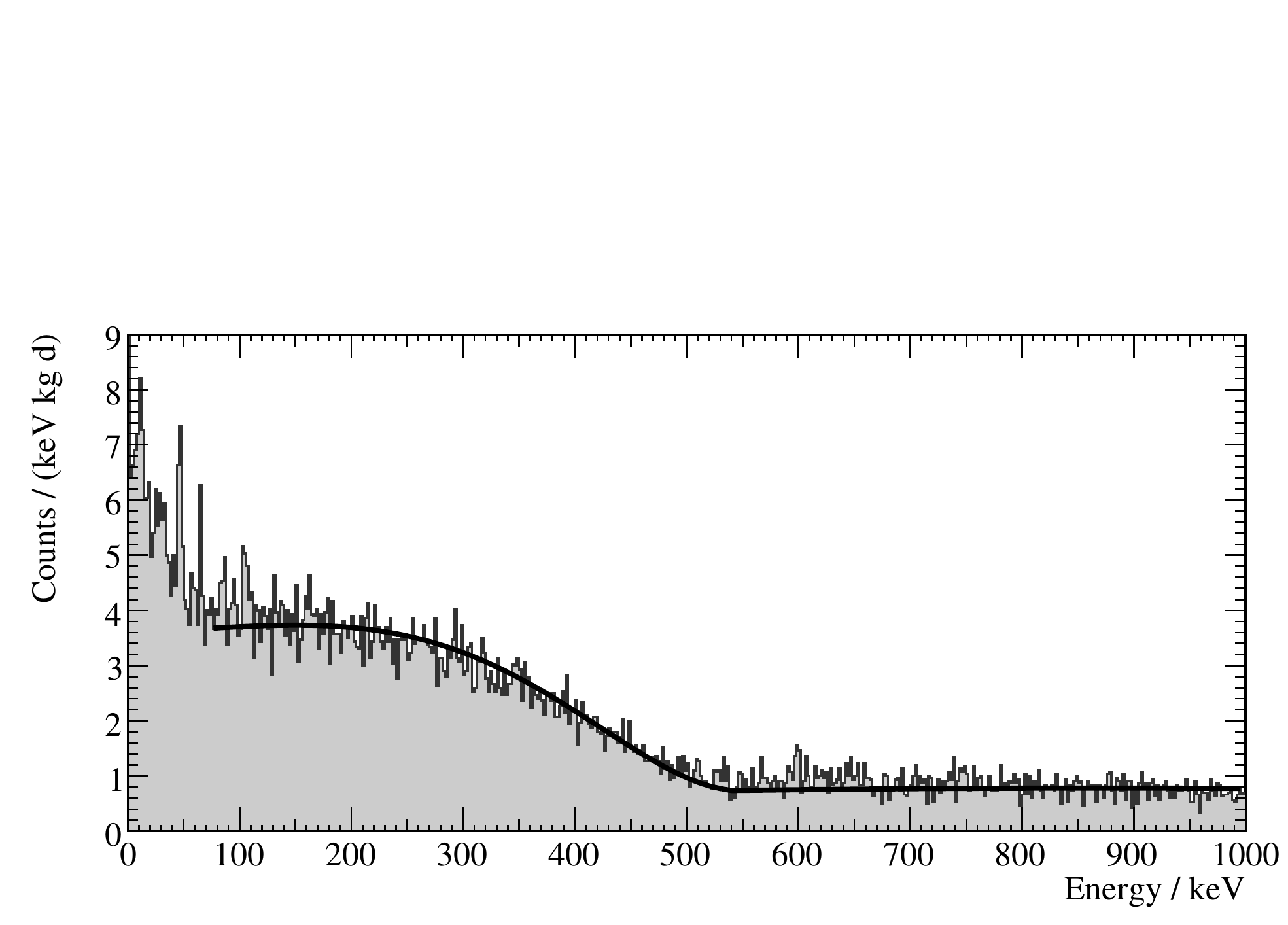}\end{center}
\vspace{-5mm}
\caption{Background spectrum observed with one crystal ($m=306.8\1{g}$) from an exposure to background radiation of $15.00\1{kg\,d}$ early 2004 (detector \textsc{Daisy}, run~28). The black line is a fit above $80\1{keV}$ of a parametrization of the $\n{{}^{90}Sr}$/$\n{{}^{90}Y}$ beta spectrum to the data.}
\label{fig:daisystrontium}\end{figure}

\begin{figure}[htb]
\begin{center}\includegraphics[angle=90,width=1\columnwidth,clip,trim=0 0 115 0]{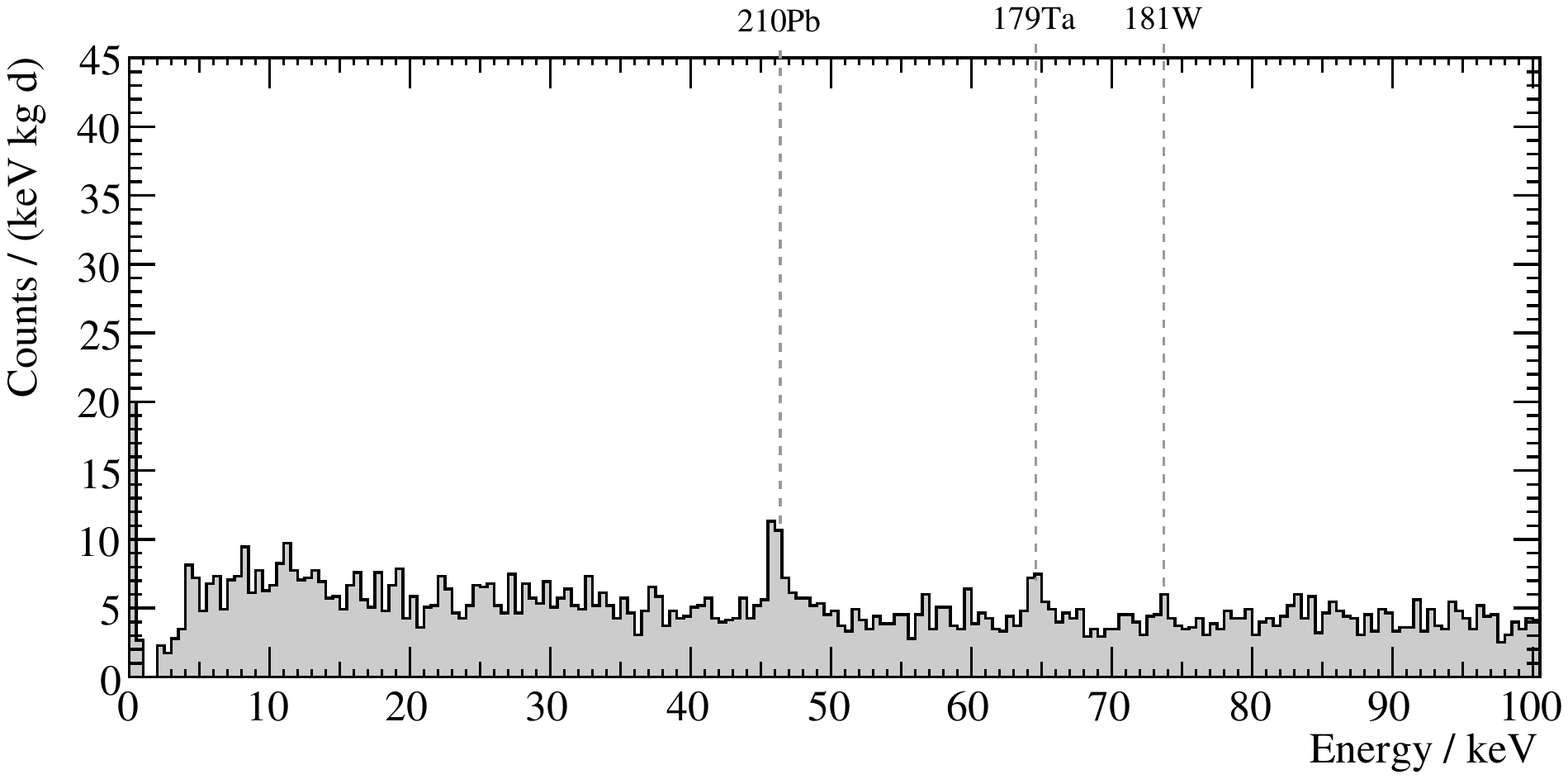}\end{center}
\vspace{-5mm}
\caption{Zoom on the energy region below $100\1{keV}$ of the spectrum shown in figure~\ref{fig:daisystrontium}.}
\label{fig:daisyrun28c}\end{figure}

Strontium belongs to the same chemical group as calcium and can thus easily be incorporated in our $\n{CaWO_4}$ crystals in the Czochralski growth process. Indeed this is the case, as can be seen in figure~\ref{fig:daisystrontium} where we show the spectrum of one crystal up to $1\1{MeV}$. We observe a continuous background which can be attributed to the beta decay of $\n{{}^{90}Sr}$ (with an endpoint energy of $546\1{keV}$) together with the subsequent beta decay of $\n{{}^{90}Y}$ ($2282\1{keV}$ endpoint energy). 

A fit above $80\1{keV}$ to a parametrization of the expected beta spectrum from $\n{{}^{90}Sr}$ in equilibrium with $\n{{}^{90}Y}$~\cite{kossert2009} can be seen to describe the data in this energy range. From the fit we deduce an activity of $(4.37\pm0.15)\1{mBq}$ of $\n{{}^{90}Sr}$ for this crystal. In addition to this common and continuous background, the three crystals show quite different features in the energy range below $100\1{keV}$.

\subsection{Lead}

Isotopes from natural decay chains are present in the crystals and can be identified by their alpha decays~\cite{cozzini2004}. A step in the natural decay of $\n{{}^{238}U}$ is $\n{{}^{210}Pb}$, which beta-decays into $\n{{}^{210}Bi}$ with an endpoint energy of $63.5\1{keV}$. In $84\percent$ of these decays a $46.5\1{keV}$ gamma is emitted. If the decays happen in the vicinity of our crystals, we observe a line at $46.5\1{keV}$ (see figure~\ref{fig:daisyrun28c}), which reflects an external activity of $(31\pm5)\1{\mu Bq}$. If, on the other hand, $\n{{}^{210}Pb}$ is a contamination intrinsic to the crystal, our calorimetric measurement gives the energy of the gamma plus that of the emitted electron. Hence, we observe a beta spectrum starting at $46.5\1{keV}$ that extends up to $63.5\1{keV}$, as can prominently be seen in figure~\ref{fig:verenarun30c}. The activity of this decay due to sources both internal and external to the crystal is much higher, namely $(672\pm28)\1{\mu Bq}$. Figure~\ref{fig:zorarun30c} also shows this lead feature for the third crystal with an activity of $(241\pm25)\1{\mu Bq}$.

\subsection{Actinium}

A similar case is that of $\n{{}^{227}Ac}$, a step in the natural decay of $\n{{}^{235}U}$. $\n{{}^{227}Ac}$ beta-decays with an endpoint energy of $44.8\1{keV}$ into $\n{{}^{227}Th}$, where two excited levels may lead to the emission of $24.5\1{keV}$ and $9.3\1{keV}$ gammas. This results in two beta spectra, one starting at $9.3\1{keV}$, and one at $24.5\1{keV}$, as can prominently be seen in figure~\ref{fig:zorarun30c} but also in figure~\ref{fig:verenarun30c}. This contamination is not observable in the spectrum shown in figure~\ref{fig:daisyrun28c}, which indicates that this detector contains less impurities from the decay of $\n{{}^{235}U}$.

\subsection{Activated Tungsten}

If the raw material or the crystals themselves are exposed to cosmic radiation, the tungsten in the $\n{CaWO_4}$ crystals can be activated. A possible channel is $\n{{}^{182}W} (p,\alpha) \n{{}^{179}Ta}$. $\n{{}^{179}Ta}$ then decays via electron capture with a half-life of 1.8 years into $\n{{}^{179}Hf}$. The energy signature of this decay in our crystals is the binding energy of the captured electron, mostly from the K-shell, which is $E_{\n{K,Hf}}=65.4\1{keV}$. The activation via $\n{{}^{183}W} (p,t)$ results in $\n{{}^{181}W}$. This decays with a half-life of 121 days via electron capture into an excited $\n{{}^{181}Ta}$ nucleus, which de-excites with emission of a $6.2\1{keV}$ gamma. Hence, the dominant energy signature of this decay is the binding energy of a tantalum K-shell electron, $E_{\n{K,Ta}}=67.4\1{keV}$, plus the energy of the gamma, $6.2\1{keV}$, adding up to $73.7\1{keV}$.

\begin{figure}[htb]
\begin{center}\includegraphics[angle=90,width=1\columnwidth,clip,trim=0 50 132 0]{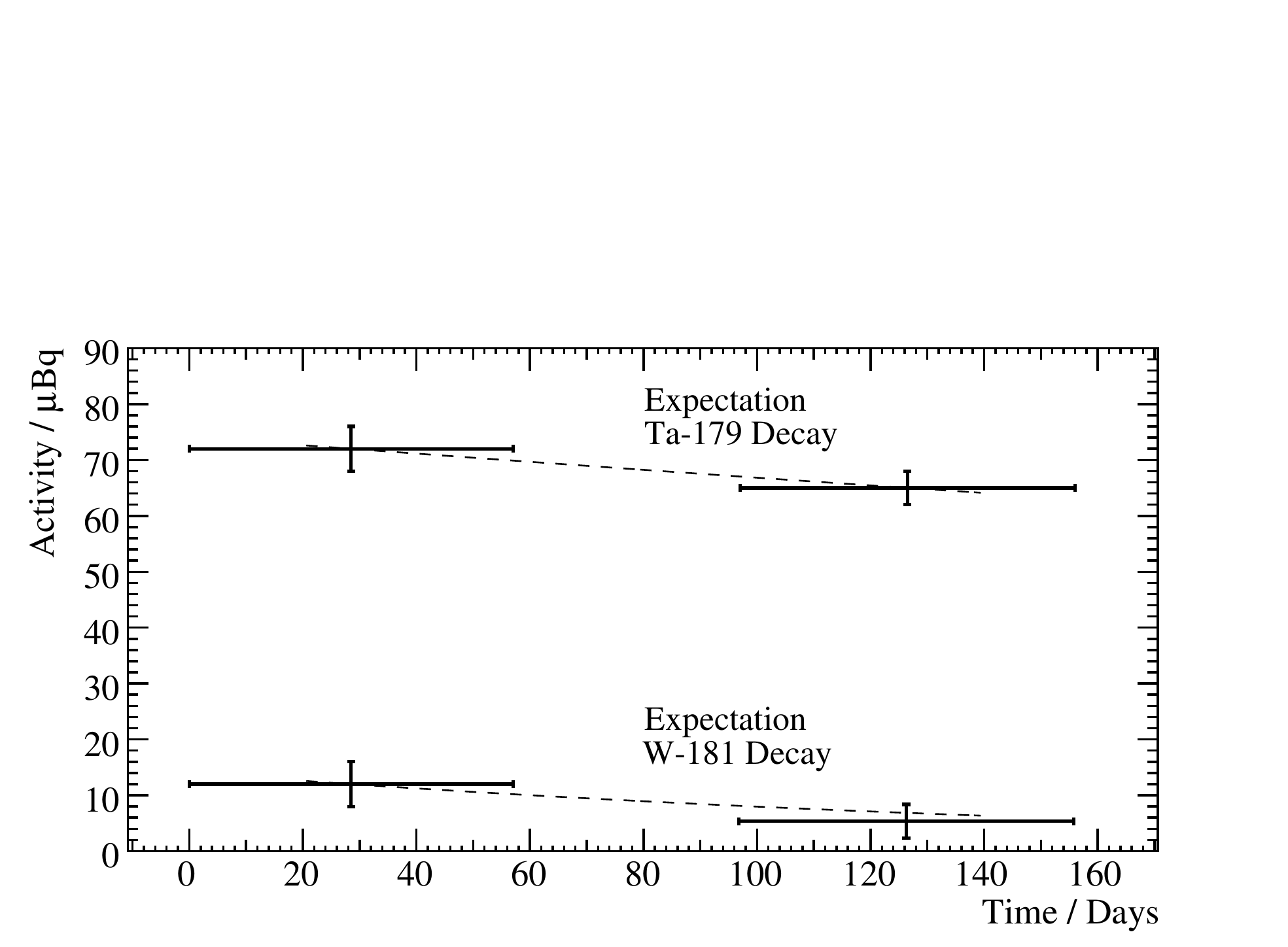}\end{center}
\vspace{-5mm}
\caption{Two measurements of the $\n{{}^{181}W}$ and $\n{{}^{179}Ta}$ activities in one crystal during two runs with similar exposure to background radiation. The expectations from decays with $T_{1/2}=121\1{d}$ and $T_{1/2}=665\1{d}$ are drawn as dashed lines and seen to be consistent with the measurement. Error bars are uncertainties of the fit to the spectral lines (each containing tens of counts) and the running time of the respective measurements (detector \textsc{Daisy}, run~27 and run~28).}
\label{fig:activated}\end{figure}

Lines at these energies can be seen in all spectra. For the crystal of which the spectrum is shown in figure~\ref{fig:daisyrun28c}, a measurement with a similar exposure to background radiation, about three months earlier, is available~\cite{majorovits2006}. This allows us to extract the $\n{{}^{179}Ta}$ and $\n{{}^{181}W}$ activities from two measurements. A comparison to the expected decay times is shown in figure~\ref{fig:activated}, confirming the origin of these lines.

\subsection{Activated Calcium}

\begin{figure}[htb]
\begin{center}\includegraphics[angle=90,width=1\columnwidth,clip,trim=0 0 115 0]{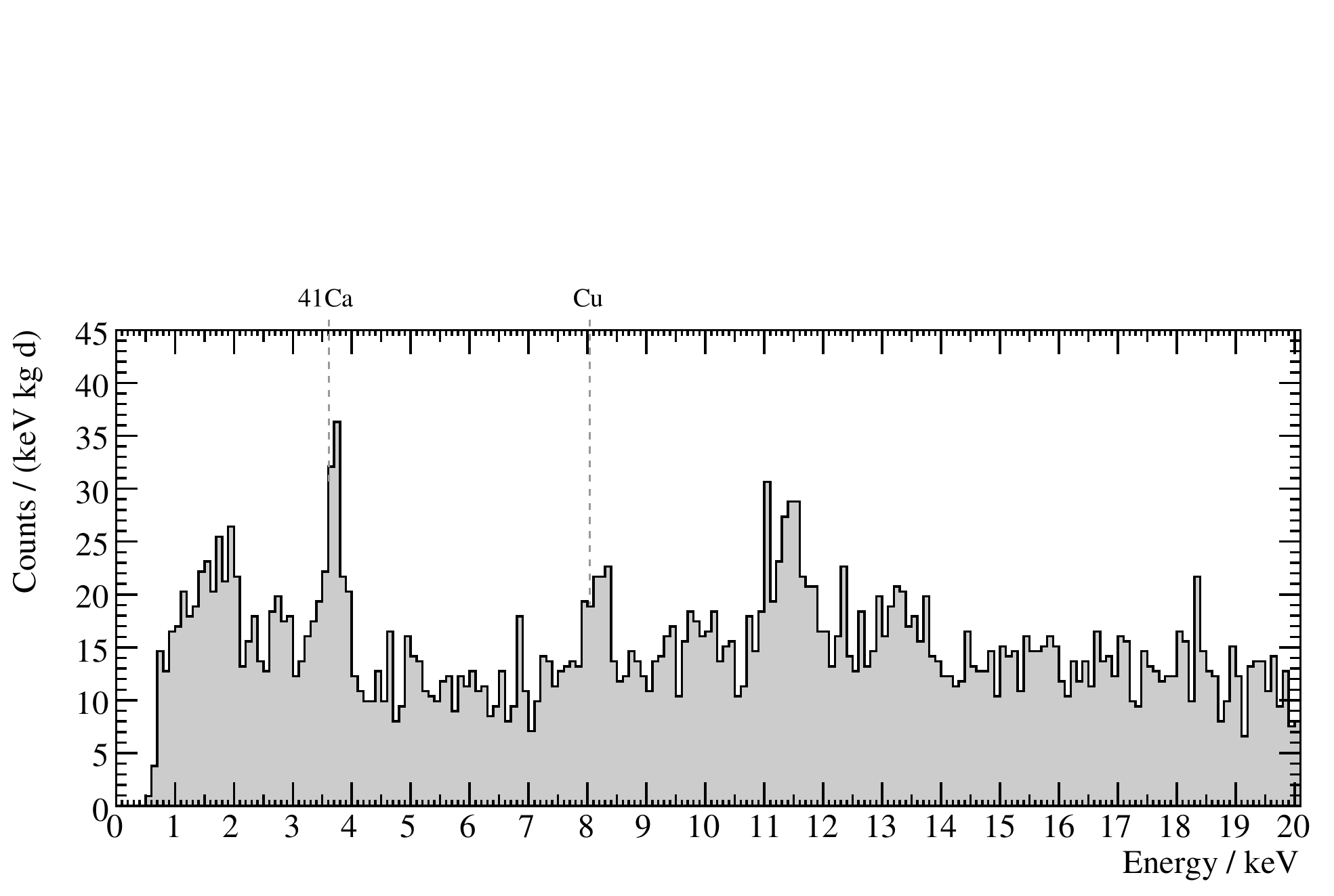}\end{center}
\vspace{-5mm}
\caption{Low-energy part of the spectrum shown in figure~\ref{fig:verenarun30c}. In addition to the lines from $\n{{}^{41}Ca}$ and copper fluorescence there is a line at $\sim11.5\1{keV}$ which so far could not be identified in a consistent way (detector \textsc{Verena}, run~30).}
\label{fig:verenarun30m}\end{figure}

\begin{figure}[htb]
\begin{center}\includegraphics[angle=90,width=1\columnwidth,clip,trim=0 0 120 0]{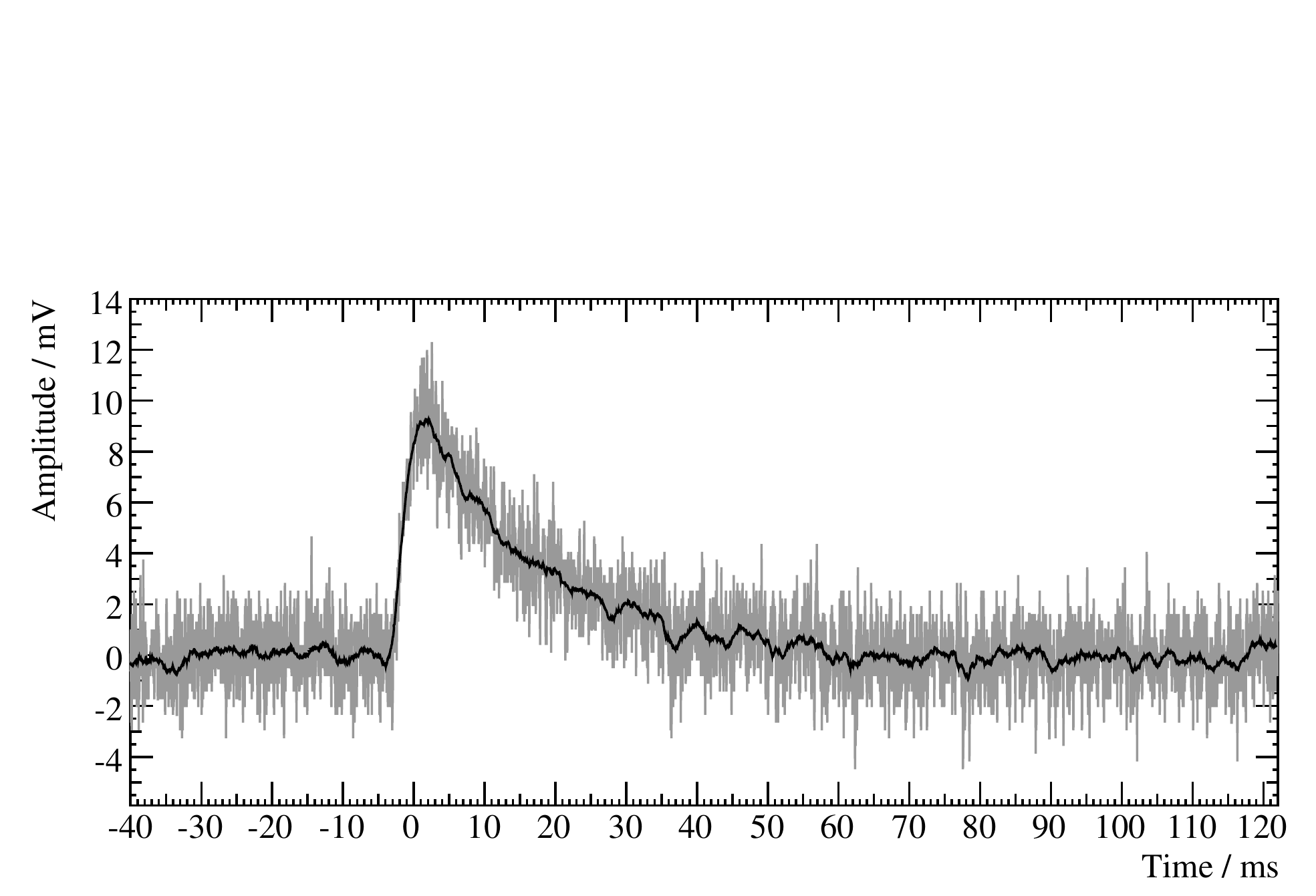}\end{center}
\vspace{-5mm}
\caption{A typical pulse with an energy of $3.7\1{keV}$ (grey), and its 50-sample ($2\1{ms}$) moving average (black).}
\label{fig:ca41event}\end{figure}

Capture of thermal neutrons can activate calcium present in the $\n{CaWO_4}$ crystals. $\n{{}^{41}Ca}$ has a half-life of $10^5$ years and decays via electron-capture to $\n{{}^{41}K}$. The K-shell binding energy of potassium is only $3.61\1{keV}$, but our detectors are capable of detecting such low energies. Figure~\ref{fig:verenarun30m} shows again the spectrum of figure~\ref{fig:verenarun30c} but for lower energies and with a finer binning. We observe a prominent line at $(3.71\pm0.02)\1{keV}$ (mean and error from the fit of a Gaussian). This is only $3\percent$ above the expected energy from $\n{{}^{41}Ca}$, which we therefore attribute to this line. The width of this line is only $(130\pm19)\1{eV}$, consistent with the calibration. For illustration, figure~\ref{fig:ca41event} shows a typical pulse at this low energy, still clearly visible above baseline noise.

From the observed activity of $(26\pm4)\1{\mu Bq}$ we can readily calculate that there are $(1.4\pm0.2)\times10^8$ $\n{{}^{41}Ca}$ atoms present in the crystal, only a fraction of $(2.2\pm0.3)\times10^{-16}$ of all calcium atoms. This is an order of magnitude more accurate than other measurements of this fraction which are typically carried out by accelerator mass spectrometry (AMS)~\cite{merchel2009,fink1990,henning1987}. Our present sensitivity is in the $10^{-17}$ range even with $\n{CaWO_4}$ crystals that are not optimized for this purpose. This demonstrates the extreme sensitivity of our detectors for rare decays and opens new possibilities for radioactive dating using $\n{{}^{41}Ca}$~\cite{raisbeck1979}. In fact, having an in situ calibration line at these low energies is an advantage for the direct Dark Matter search, since it independently confirms our energy calibration procedure in this otherwise inaccessible region.

\subsection{Copper Fluorescence}\label{sec:copper}

Another line, seen at $(8.17\pm0.04)\1{keV}$ in figure~\ref{fig:verenarun30m}, can be attributed to copper fluorescence. This originates from gammas or electrons that eject a photoelectron from the copper of the cryostat that surrounds the detectors, resulting in a copper X-ray absorbed in our crystals. Our energy calibration puts the line only $2\percent$ higher than the expected $8.04\1{keV}$ from the literature. This is consistent with the slight energy overestimate seen when identifying the $\n{{}^{41}Ca}$ spectral line, again confirming its origin.

\subsection{Lutetium}

\begin{figure}[htb]
\begin{center}\includegraphics[angle=90,width=1\columnwidth,clip,trim=0 0 115 0]{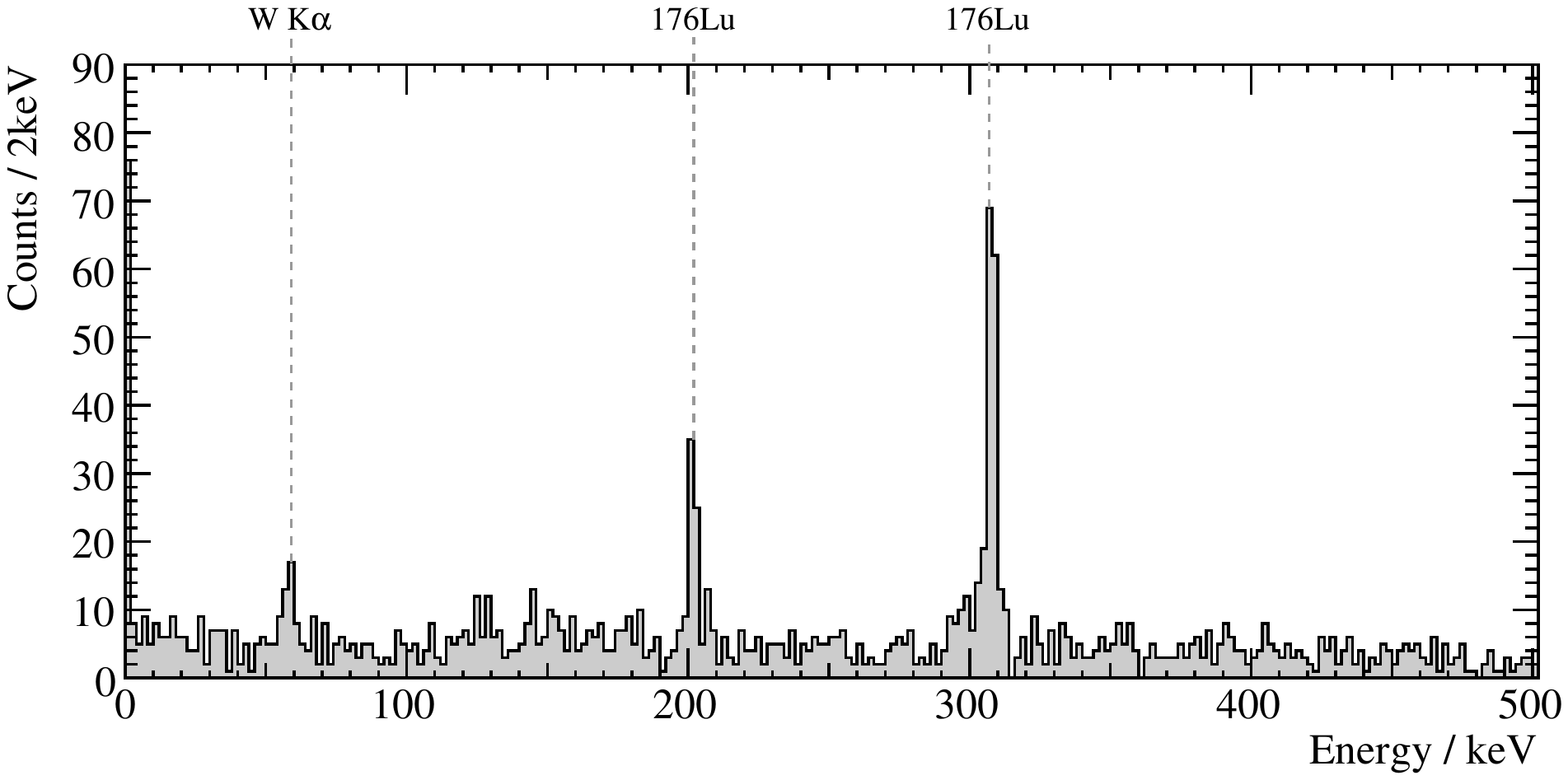}\end{center}
\vspace{-5mm}
\caption{Spectrum of coincident events. To increase the counting statistics, the spectra of the two individual detectors were added to produce this figure. Without the coincidence requirement, the added spectrum would have a count rate of $\order (500)$ counts per $2\1{keV}$ bin (detectors \textsc{Verena} and \textsc{Zora}, run~30).}
\label{fig:coincidentb}\end{figure}

All single crystal spectra above $100\1{keV}$ are rather featureless. Yet, two crystals were operated simultaneously and in close vicinity during the commissioning of the setup. This allows us to search for coincident events, of which a spectrum is shown in figure~\ref{fig:coincidentb}. 

\begin{figure}[htb]
\begin{center}\includegraphics[width=.8\columnwidth,clip,trim=0 0 0 60]{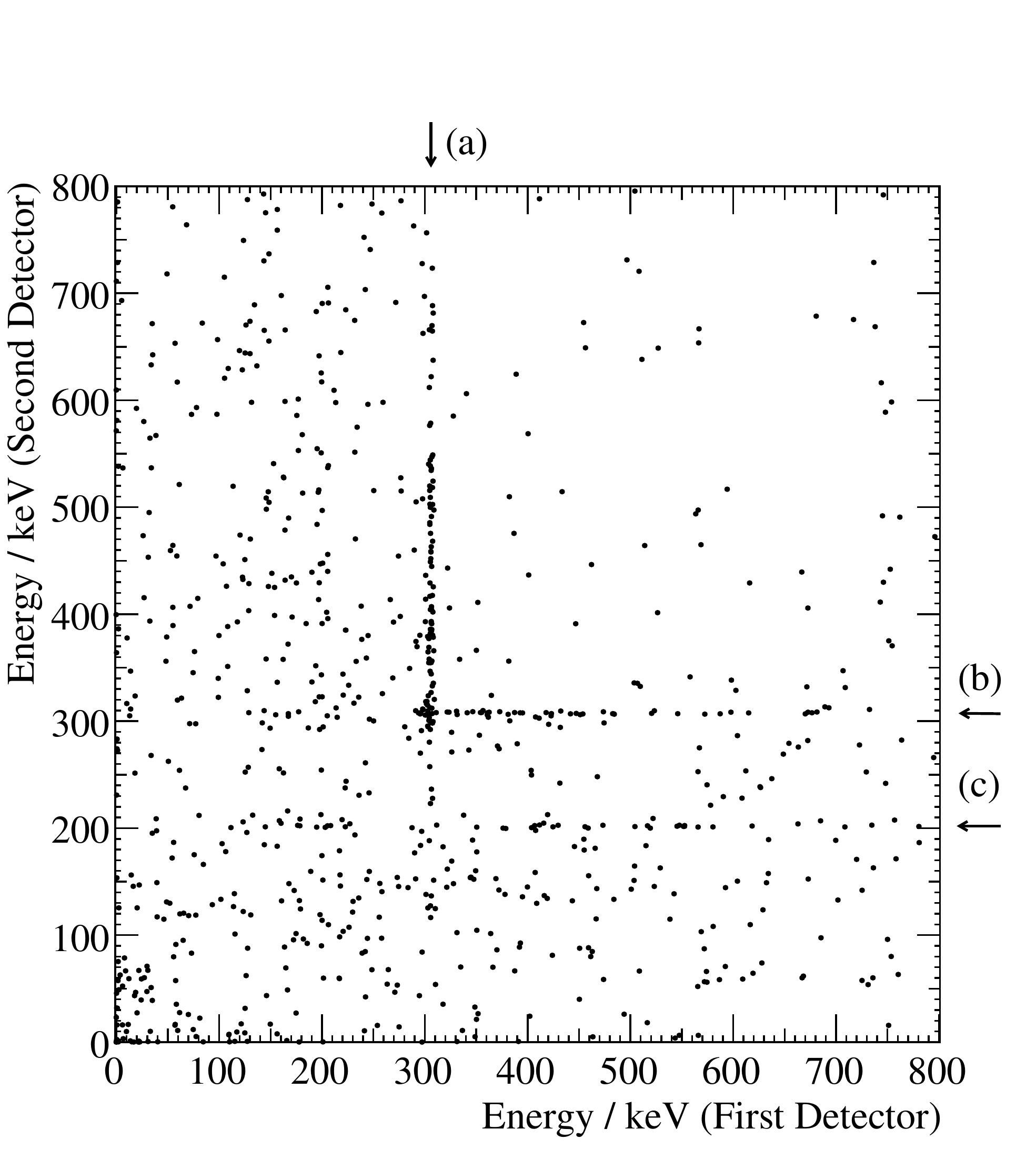}\end{center}
\vspace{-5mm}
\caption{Scatter plot of coincident events in two detectors operated during commissioning. The vertical line~(a) going up from $290\1{keV}$ is due to lutetium decaying in the second detector, in coincidence with the $307\1{keV}$ gamma escaped to the first detector. The horizontal line~(b) is due to the same process with detectors reversed, at $307\1{keV}$ in the second detector and starting at $290\1{keV}$ in the first one. At~(c) we observe a hint of a horizontal line starting at $395\1{keV}$ due to lutetium decaying in the first detector with the energy of an escaped $202\1{keV}$ gamma in the second detector.}
\label{fig:zoravsverenab}\end{figure}

\begin{figure}[htb]
\begin{center}\includegraphics[angle=90,width=1\columnwidth,clip,trim=0 50 132 0]{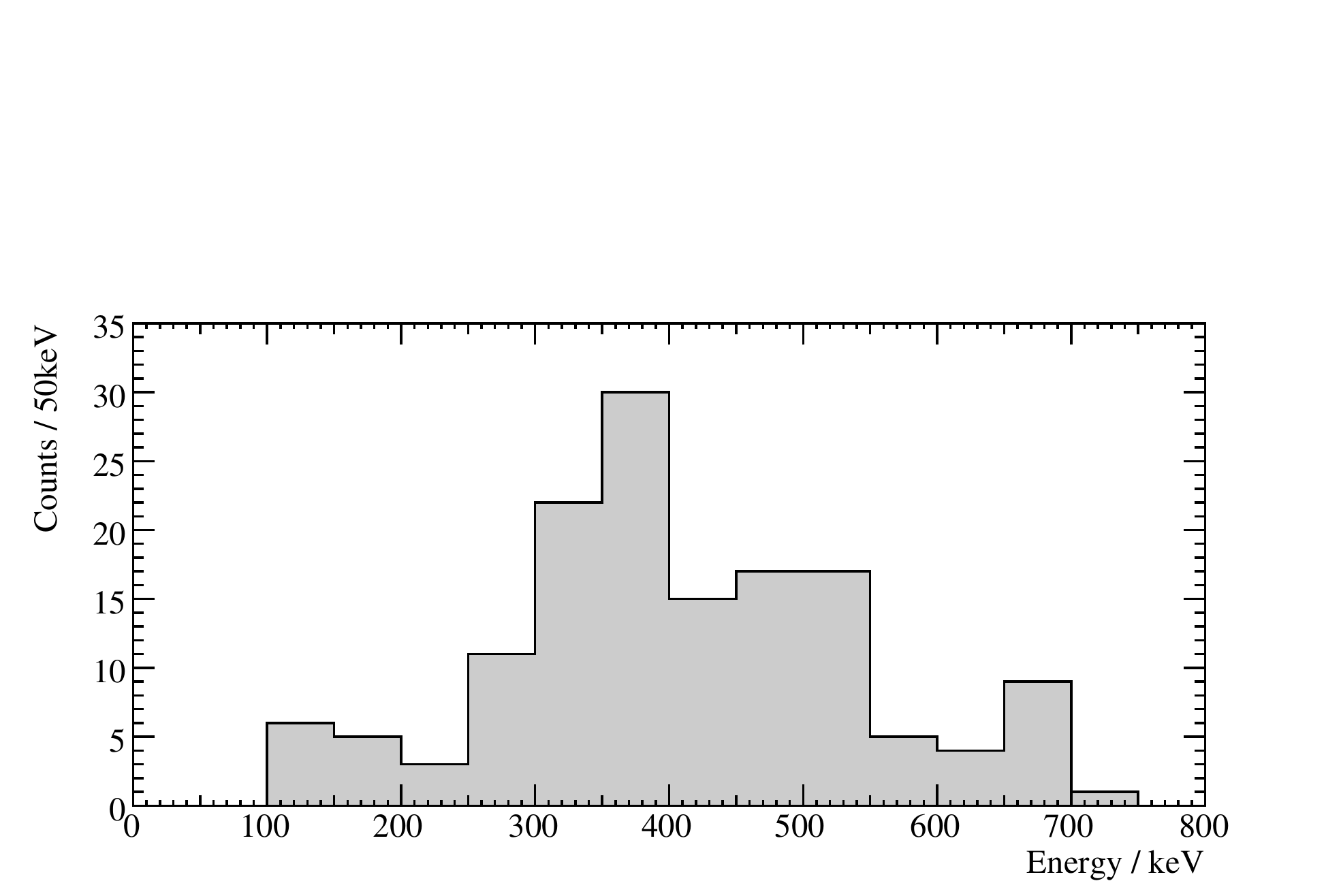}\end{center}
\vspace{-5mm}
\caption{Histogram of all coincident events in one detector that appear at $(307\pm3)\1{keV}$ in the respective other detector: The $\n{{}^{176}Lu}$ beta spectrum starting at $290\1{keV}$ is visible despite the low statistics.}
\label{fig:beta}\end{figure}

At $59\1{keV}$, tungsten $\n{K_{\alpha}}$ escape events can be seen. In addition, two lines at $202\1{keV}$ and $307\1{keV}$ become very prominent, and we attribute those to a contamination with lutetium intrinsic to the crystals. $\n{{}^{176}Lu}$ beta-decays to $\n{{}^{176}Hf}$ with a total decay energy of $1191.7\1{keV}$. Each decay produces a gamma cascade with gamma energies of $306.8\1{keV}$, $201.8\1{keV}$ and $88.3\1{keV}$. In particular the higher energy gammas may escape from one crystal and hit the other. Hence, for a coincident $306.8\1{keV}$ event observed in one detector, the other crystal will mostly absorb the other two gammas, plus the energy of the beta electron. This explains the structure seen in the scatter plot of figure~\ref{fig:zoravsverenab}, showing the energies of coincident events in each detector. We observe a beta spectrum starting at $201.8\1{keV}+88.3\1{keV}=290.1\1{keV}$ for most events at $306.8\1{keV}$ in the other detector (lines marked~(a) and~(b) in figure~\ref{fig:zoravsverenab}). Figure~\ref{fig:beta} shows the energy spectrum of these events in one detector, tagged by an energy of $306.8\1{keV}$ in the other one, in agreement with the expectation from the lutetium beta spectrum.

\section{Conclusions}

The excellent energy resolution of less than $1\1{keV}$ over the full energy range used in the Dark Matter search~\cite{angloher2005,angloher2009} allows us to identify a variety of different background sources in the CRESST experiment. Since the CRESST experiment can discriminate nuclear recoil signal from electron recoil background based on the light output, the presence of these features does not compromise the search for Dark Matter, yet allows an audit of the performance of the experiment. The energy resolution ($1\sigma$) is $130\1{eV}$ at $3.6\1{keV}$, and the observed sharpness of the lines proves the high stability of the detectors during the background measurements running for several months. The observed line from activated calcium provides an independent in situ energy calibration even below the energy range used for the Dark Matter search, and in addition proves the high sensitivity of the experiment to rare decays.

\section{Acknowledgments}

R.~F.~L. acknowledges useful discussions with G.~Korschinek on $\n{{}^{41}Ca}$. This work was partially supported by funds of the DFG (SFB 375 and Transregio 27 ``Neutrinos and Beyond''), the Munich Cluster of Excellence (``Origin and Structure of the Universe"), the EU networks for Cryogenic Detectors (ERB-FMRXCT980167) and for Applied Cryogenic Detectors (HPRN-CT2002-00322), and the Maier-Leibnitz-Laboratorium (Garching). Support was provided by the Science and Technology Facilities Council. 


\end{document}